\documentclass[prb,preprint,eqsecnum]{revtex4}%
\usepackage{amsmath}
\usepackage{graphicx}
\usepackage{amssymb}
\usepackage{amsfonts}%
\begin{document}
\title{New series representation for Madelung constant}
\author{Sandeep Tyagi}
\email{satst27@pitt.edu}
\affiliation{Department of Chemical and Petroleum Engineering, University of Pittsburgh,
Pittsburgh, Pennsylvania 15261}

\begin{abstract}
A new series for the Madelung constant $M$ is derived on the basis of a
representation given by R.~Crandall (1999). We are able to write $M = C + S$,
where $S$ is a rapidly convergent series, and the constant $C$
is fundamental:
$$C = -\frac{1}{8}-\frac{\ln  2 }{4\pi}-\frac{4\pi}
{3}+\frac{1}{2\sqrt{2}}+\frac{\Gamma\left(  \frac{1}{8}\right)  \Gamma\left(
\frac{3}{8}\right)  }{\pi^{3/2}\sqrt{2}}  \ \approx -1.747564594\dots.$$
The remarkable result is that even if the $S$
term be discarded, this constant $C$ gives 10 good decimal places of $M$.
This result advances the state of the art in the discovery of what Crandall has termed ``close calls" to an exact Madelung
evaluation.  We indicate related identities and how this fundamental 10-digit accuracy
might be further enhanced.
\end{abstract}
\maketitle

\section{Introduction}

The Madelung constant has fascinated mathematicians and physicists alike over the
past century. It is physically related to the electrostatic interaction
of a sodium ion with all other ions in a perfect NaCl crystal. The Madelung
constant, $M$, can be written as%
\begin{equation}
M=\sum_{m,n,p}^{\prime}\frac{\left(  -1\right)  ^{m+n+p}}{\sqrt{m^{2}%
+n^{2}+p^{2}}}, \label{e1}%
\end{equation}
where the sum over integers $m,n$ and $p$ runs from $-\infty$ to
$+\infty$. A prime over the summation sign indicates that the term
corresponding to the zero vector $(m,n,p) = (0,0,0)$ be avoided. The
sum in eq.(\ref{e1}) is 
only conditionally convergent. One way to proceed is to interpret
eq.(\ref{e1}) as a sum over expanding cubes (Borwein et. al 1985).
Another approach is to interpret $M$ as the analytic continuation of
a sum---with $(m^2 + n^2 + p^2)^{-s}$ appearing with $\Re(s) > 3/2$---to 
a definite value at $s = 1/2$.

It has been an aim of many researchers to recast such lattice sums in
terms of functions that decay exponentially fast for large $r$ $=\sqrt
{m^{2}+n^{2}+p^{2}}.$ Equation (\ref{e1}) can be evaluated using Ewald method
(Ewald 1920). The Ewald expansion certainly enjoys rapid decay, but error functions
need be evaluated and this can be computationally problematic.
 Important work on Madelung constant and lattice sums
has been carried out by Hautot (1975), Zucker (1976), Glasser and Zucker
(1980), Borwein et. al (1985) and Crandall and Buhler (1987). 
Many of these historical treatments involve rapidly decaying sums of {\it elementary}
functions.
 A fascinating account of Madelung constant, together with some new representations
 (such as a finite-domain integral representation), has been
given by Crandall (1999).  The present work uses that reference essentially
as a starting point in a new quest to accelerate convergence.

\section{Madelung constant}

We start with the identity%

\begin{align}
\sum_{k=-\infty}^{\infty}\frac{1}{\left[  \left(  x+k\right)  ^{2}%
+r^{2}\right]  ^{\frac{1}{2}+\nu}}  &  =\frac{\sqrt{\pi}}{\Gamma\left(
\nu+\frac{1}{2}\right)  }\left\{  \frac{\Gamma\left(  \nu\right)  }{r^{2\nu}%
}+4\left(  \frac{\pi}{r}\right)  ^{\nu}\right. \label{e2}\\
&  \left.  \times \sum_{l=1}^{\infty}l^{\nu}K_{\nu}\left(  2\pi lr\right)
\cos\left(  2\pi lx\right)  \right\}  .\nonumber
\end{align}
This identity can be found in Sperb (1996). A special case of the
above identity is also found in Glasser and Zucker (1980). An easy proof may
of the identity may be obtained with an application of the Possion summation
theorem to the integral (Gradshteyn and Ryzhik 1965)%
\begin{equation}
K_{\nu}\left(  \lambda x\right)  =\frac{\Gamma\left(  \nu+\frac{1}{2}\right)
\left(2 \lambda\right)^{\nu}}{\sqrt{\pi}x^{\nu}}\int_{0}^{\infty}dk\frac{\cos\left(  kx\right)  }{\left(
k^{2}+\lambda^{2}\right)  ^{\nu+\frac{1}{2}}}. \label{e3}%
\end{equation}
We note an important aspect of the identity in eq. (\ref{e2}). With the
help of this identity we can relate a summation in $2\nu+1$ dimensional space
to a summation in $2\nu$ dimensional space. In fact taking $\nu$ $=0$ one can
easily obtain a generalization of Lekner's (1998) work on Coulomb sums for a
triclinic cell.

Putting $x=0,\nu=1/2$, $r=\sqrt{m^{2}+n^{2}+p^{2}}$ and summing over integers
$m,n$ and $p$ we obtain
\begin{align}
S  &  =\sum_{k=-\infty}^{\infty}\sum_{m,n,p}^{\prime}\frac{\left(  -1\right)
^{m+n+p}}{k^{2}+m^{2}+n^{2}+p^{2}}\label{e4}\\
&  =\pi\sum_{m,n,p}^{\prime}\frac{\left(  -1\right)  ^{m+n+p}}{\sqrt
{m^{2}+n^{2}+p^{2}}}+4\pi\sum_{m,n,p}^{\prime}\frac{\left(  -1\right)
^{m+n+p}}{\sqrt{m^{2}+n^{2}+p^{2}}\left[  \exp\left(  2\pi\sqrt{m^{2}%
+n^{2}+p^{2}}\right)  -1\right]  },\nonumber
\end{align}
where we have used
\begin{equation}
K_{1/2}\left(  x\right)  =\sqrt{\frac{\pi}{2x}}\exp\left(  -x\right)  .
\label{e5}%
\end{equation}
Now, the second sum on the lhs in eq.(\ref{e4}) can be obtained in a closed form,%

\begin{align}
S  &  =\lim_{\alpha\rightarrow0}\sum_{k=-\infty}^{\infty}\sum_{m,n,p}^{\prime
}\frac{\left(  -1\right)  ^{m+n+p}}{k^{2}+m^{2}+n^{2}+p^{2}+\alpha^{2}%
}\label{e6}\\
&  =\lim_{\alpha\rightarrow0}\left(  \sum_{k,m,n,p}\frac{\left(  -1\right)
^{m+n+p}}{k^{2}+m^{2}+n^{2}+p^{2}+\alpha^{2}}-\sum_{k}\frac{1}{k^{2}%
+\alpha^{2}}\right) \nonumber\\
&  =\sum_{k,m,n,p}^{\prime}\frac{\left(  -1\right)  ^{m+n+p}}{k^{2}%
+m^{2}+n^{2}+p^{2}}+\lim_{\alpha\rightarrow0}\left(  \frac{1}{\alpha^{2}}%
-\sum_{k}\frac{1}{k^{2}+\alpha^{2}}\right)  .\nonumber
\end{align}
The first sum on the rhs of eq. (\ref{e6}) may be evaluated with help of results given
in Zucker (1984):%
\begin{equation}
\sum_{k,m,n,p}^{\prime}\frac{\left(  -1\right)  ^{m+n+p}}{k^{2}+m^{2}%
+n^{2}+p^{2}}=-\frac{\pi}{2}-\ln  2 , \label{e7}%
\end{equation}
while the second part is simply%
\begin{equation}
\lim_{\alpha\rightarrow0}\left(  \frac{1}{\alpha^{2}}-\sum_{k}\frac{1}%
{k^{2}+\alpha^{2}}\right)  =\lim_{\alpha\rightarrow0}\left(  \frac{1}%
{\alpha^{2}}-\frac{\pi\coth\left(  \pi\alpha\right)  }{\alpha}\right)
=-\frac{\pi^{2}}{3}. \label{e8}%
\end{equation}
Thus we obtain
\begin{align}
M  &  =-\frac{1}{2}-\frac{\ln 2  }{\pi}-\frac{\pi}{3}%
\label{e9}\\
&  -2\sum_{m,n,p}^{\prime}\frac{\left(  -1\right)  ^{m+n+p}}{\sqrt{m^{2}%
+n^{2}+p^{2}}\left[  \exp\left(  2\pi\sqrt{m^{2}+n^{2}+p^{2}}\right)
-1\right]  }.\nonumber
\end{align}
Equation (\ref{e9}) is one of the important results of this paper. We
will now derive an alternate expression for $M$ again starting from the basic
identity in eq.(\ref{e2}). This time we substitute $x=1/2,$ which
summing over $m,n$ and $p$ as before leads to
\begin{align}
\sum_{k=-\infty}^{\infty}\sum_{m,n,p}^{\prime}\frac{\left(  -1\right)
^{m+n+p}}{\left(  k+\frac{1}{2}\right)  ^{2}+m^{2}+n^{2}+p^{2}}  &  =\pi
\sum_{m,n,p}^{\prime}\frac{\left(  -1\right)  ^{m+n+p}}{\sqrt{m^{2}%
+n^{2}+p^{2}}}\label{e10}\\
&  -2\pi\sum_{m,n,p}^{\prime}\frac{\left(  -1\right)  ^{m+n+p}}{\sqrt
{m^{2}+n^{2}+p^{2}}\left[  \exp\left(  2\pi\sqrt{m^{2}+n^{2}+p^{2}}\right)
+1\right]  }.\nonumber
\end{align}
Again the rhs of eq. (\ref{e10}) can be obtained in a closed form. We
start with 
\begin{align}
&  \sum_{k=-\infty}^{\infty}\sum_{m,n,p}^{\prime}\frac{\left(  -1\right)
^{m+n+p}}{\left(  k+\frac{1}{2}\right)  ^{2}+m^{2}+n^{2}+p^{2}}\label{e11}\\
&  =\sum_{k,m,n,p}\frac{\left(  -1\right)  ^{m+n+p}}{\left(  k+\frac{1}%
{2}\right)  ^{2}+m^{2}+n^{2}+p^{2}}-\sum_{k=-\infty}^{\infty}\frac{1}{\left(
k+\frac{1}{2}\right)  ^{2}}.\nonumber
\end{align}
The first sum on the rhs is again be obtained from the results
contained in Zucker (1984):%
\begin{equation}
\sum_{k,m,n,p}\frac{\left(  -1\right)  ^{m+n+p}}{\left(  k+\frac{1}{2}\right)
^{2}+m^{2}+n^{2}+p^{2}}=\pi\sqrt{2}. \label{e12}%
\end{equation}
 The second sum on the rhs of Eq.(\ref{e11}) can be easily evaluated to
be $\pi^{2}.$ Thus we obtain%
\begin{equation}
M=\sqrt{2}-\pi+2\sum_{m,n,p}^{\prime}\frac{\left(  -1\right)  ^{m+n+p}}%
{\sqrt{m^{2}+n^{2}+p^{2}}\left[  \exp\left(  2\pi\sqrt{m^{2}+n^{2}+p^{2}%
}\right)  +1\right]  }. \label{e13}%
\end{equation}
 We can get a third and fast converging expression for $M$ by taking
average over $M$ from eqs.(\ref{e9}) and (\ref{e13}). We obtain%
\begin{align}
M  &  =-\frac{1}{4}-\frac{\ln   2  }{2\pi}-\frac{2\pi}{3}+\frac
{1}{\sqrt{2}}\label{e14}\\
&  -2\sum_{m,n,p}^{\prime}\frac{\left(  -1\right)  ^{m+n+p}}{\sqrt{m^{2}%
+n^{2}+p^{2}}\left[  \exp\left(  4\pi\sqrt{m^{2}+n^{2}+p^{2}}\right)
-1\right]  }.\nonumber
\end{align}
 This is not the whole story yet. We can now make use of a beautiful
expression for $M$ derived by Crandall (1999):%
\begin{align}
M  &  =-2\pi+\frac{\Gamma\left(  \frac{1}{8}\right)  \Gamma\left(  \frac{3}%
{8}\right)  \sqrt{2}}{\pi^{3/2}}\label{e15}\\
&  +2\sum_{m,n,p}^{\prime}\frac{\left(  -1\right)  ^{m+n+p}}{\sqrt{m^{2}%
+n^{2}+p^{2}}\left[  \exp\left(  4\pi\sqrt{m^{2}+n^{2}+p^{2}}\right)
+1\right]  }.\nonumber
\end{align}
 Taking the average over the two values of $M$ from Eq.(\ref{e14}) and
Eq.(\ref{e15}) we obtain%
\begin{align}
M  &  =-\frac{1}{8}-\frac{\ln  2  }{4\pi}-\frac{4\pi}{3}+\frac
{1}{2\sqrt{2}}+\frac{\Gamma\left(  \frac{1}{8}\right)  \Gamma\left(  \frac
{3}{8}\right)  }{\pi^{3/2}\sqrt{2}}\label{e16}\\
&  -2\sum_{m,n,p}^{\prime}\frac{\left(  -1\right)  ^{m+n+p}}{\sqrt{m^{2}%
+n^{2}+p^{2}}\left[  \exp\left(  8\pi\sqrt{m^{2}+n^{2}+p^{2}}\right)
-1\right]  }.\nonumber
\end{align}
This representation of $M$ is of the form $C + S$ of our Abstract.
 Even if the whole series part in Eq.(\ref{e16}) be ignored, we obtain%
\begin{align}
M  &  \approx  C := \ -\frac{1}{8}-\frac{\ln  2  }{4\pi}-\frac{4\pi}%
{3}+\frac{1}{2\sqrt{2}}+\frac{\Gamma\left(  \frac{1}{8}\right)  \Gamma\left(
\frac{3}{8}\right)  }{\pi^{3/2}\sqrt{2}}\label{e17}\\
&  =-1.747564594(7).\nonumber
\end{align}
 The value of $M$ thus obtained is correct to an astounding 10 good decimals.

The present method also leads to identities such as mentioned below.
Eliminating $M$ from Eq. (\ref{e9}) and Eq. (\ref{e13}) we obtain%
\begin{equation}
\sum_{m,n,p}^{\prime}\frac{\left(  -1\right)  ^{m+n+p}}{\sqrt{m^{2}%
+n^{2}+p^{2}}}\text{csch}\left(  2\pi\sqrt{m^{2}+n^{2}+p^{2}}\right)
=-\frac{1}{4}-\frac{\ln 2  }{2\pi}+\frac{\pi}{3}-\frac{1}%
{\sqrt{2}}. \label{e18}%
\end{equation}
Similarly combining eq. (\ref{e14}) and eq.(\ref{e15}) one obtains%
\begin{align}
\sum_{m,n,p}^{\prime}\frac{\left(  -1\right)  ^{m+n+p}}{\sqrt{m^{2}%
+n^{2}+p^{2}}}\text{csch}\left(  4\pi\sqrt{m^{2}+n^{2}+p^{2}}\right)   &
=-\frac{1}{8}-\frac{\ln 2  }{4\pi}+\frac{2\pi}{3}\label{e19}\\
&  +\frac{1}{2\sqrt{2}}-\frac{\Gamma\left(  \frac{1}{8}\right)  \Gamma\left(
\frac{3}{8}\right)  }{\pi^{3/2}\sqrt{2}}.\nonumber
\end{align}
 Equations (\ref{e18}) and (\ref{e19}) can be represented in alternate
form involving four dimensional sums. Another important identity that can be
derived with the help of eq.(\ref{e2}) and (\ref{e15}) is%
\begin{equation}
\sum_{k,m,n,p}\frac{\left(  -1\right)  ^{m+n+p}}{\left(  k+\frac{1}{2}\right)
^{2}+4\left(  m^{2}+n^{2}+p^{2}\right)  }=\frac{\Gamma\left(  \frac{1}%
{8}\right)  \Gamma\left(  \frac{3}{8}\right)  }{\sqrt{2 \pi}}. \label{e20}%
\end{equation}

\section{Conclusion}

We have given a new representation of the Madelung constant. The
representation given here is very fast. Ignoring the series altogether, one
still obtains Madelung constant correct up to ten decimal
figures. Here I note that if one could
obtain the sum%
\begin{equation}
\sum_{k,m,n,p}\frac{\left(  -1\right)  ^{m+n+p}}{\left(  k+\frac{1}{2}\right)
^{2}+16\left(  m^{2}+n^{2}+p^{2}\right)  }, \label{e21}%
\end{equation}
in a closed from, then by combining that result with the one given here,
one may obtain a representation of Madelung constant, where even dropping
the series part, one would still obtain Madelung constant correct close to 20
decimal points. A probable approach may be through theta function identities
involving q-series. However, we are not familiar with a systematic method of
converting a product of theta function into a q-series, if one exist. It is
hoped that the approach pointed out here may open up ways
for researchers to look afresh on whether the Madelung constant can be
expressed in a closed form.

\acknowledgments{I am thankful to Prof. R.~Crandall for helpful
  comments and mathematical advice.}

\end{document}